\documentclass[sigconf]{acmart}
\usepackage{url}
\usepackage{comment}
\usepackage[subrefformat=parens]{subcaption}

\setcopyright{none}              
\renewcommand\footnotetextcopyrightpermission[1]{%
  \footnotetext{%
    \copyright\ ACM 2025. This is the author's version of the work. It is posted here for your personal use. Not for redistribution. The definitive Version of Record was published in \textit{Proceedings of the 3rd International Workshop on Middleware for the Computing Continuum (Mid4CC) 2025}, \url{https://doi.org/10.1145/3774898.3778040}.%
  }%
}

\begin{document}

\title{Self-Hosted WebAssembly Runtime for Runtime-Neutral\\Checkpoint/Restore in Edge–Cloud Continuum}


\author{Yuki Nakata}
\affiliation{
  \institution{SAKURA internet Inc. / Future University Hakodate}
  \state{Hokkaido}
  \country{Japan}
}
\email{y-nakata@sakura.ad.jp}

\author{Katsuya Matsubara}
\affiliation{
  \institution{Future University Hakodate}
  \state{Hokkaido}
  \country{Japan}
}
\email{matsu@fun.ac.jp}

\begin{abstract}
  Checkpoint/restore (C/R) in the Edge–Cloud Continuum, where edge servers and cloud servers are integrated to combine their respective computational characteristics, enables application live migration for load balancing and service continuity that adapts to user mobility.
  WebAssembly (Wasm) provides architecture neutrality, allowing bytecode execution across diverse platforms.
  Furthermore, by leveraging runtime implementations specialized for particular environments and performance optimizations such as JIT and AOT, Wasm bytecode can be executed efficiently by using the runtime best suited to each environment.
  However, the diversity of runtimes and performance optimizations results in heterogeneous representations of application execution states, making it difficult to apply C/R across multiple runtimes.
  To address this heterogeneity, we propose a C/R method that employs a self-hosted Wasm runtime compiled into Wasm itself.
  By checkpointing and restoring application execution states within the self-hosted Wasm runtime, the representation of execution states can be unified.
  In addition, to mitigate the overhead of duplicated runtime executions, we also investigate optimization techniques for Wasm runtimes designed to be self-hosted.
  Our evaluation results demonstrate that C/R with a self-hosted Wasm runtime can eliminate runtime differences and enable efficient live migration with a minimal execution state.
\end{abstract}



\maketitle
\pagestyle{plain}                

\section{Introduction}
In the Edge–Cloud continuum \cite{edge-cloud-continuum}, which is a distributed computing environment that integrates edge and cloud resources, a runtime that can run applications on diverse CPU architectures without recompilation and support live migration is vital.
Edge servers provide low-latency processing and efficient data access close to end users, while cloud servers achieve high-performance processing of large-scale data, thus exhibiting different processing characteristics.
In addition, edge servers often use ARM CPUs whereas cloud servers commonly utilize x86-64 CPUs, so hardware specifications differ between servers \cite{hetero-architecture-fog-edge-computing}, requiring separate applications to be compiled for each environment and the consideration of architectural differences between servers.
Therefore, enabling applications to run on diverse architectures without recompilation and to be moved among them through live migration enables scenarios such as offloading compute-intensive tasks from devices or edge servers to cloud servers, as well as handing off computation among edge servers to follow user mobility.

WebAssembly (Wasm), a virtual instruction set architecture, is useful as an architecture-neutral application execution platform for the Edge–Cloud continuum \cite{wasm-edge-cloud-continuum}.
Wasm allows programs written in various languages to be compiled into Wasm bytecode and executed on a virtual machine (VM) provided by a Wasm runtime.
Wasm can execute software on diverse architectures and platforms, such as the edge and the cloud, without recompilation.

The use of suitable Wasm runtimes on each platform, along with checkpoint/restore (C/R) of Wasm bytecode among them, enables efficient Wasm bytecode execution and live migration between the edge and the cloud.
Many Wasm runtimes have been developed for different environments and use cases, including devices, the edge, and the cloud \cite{wasm-runtimes}.
These runtimes implement performance optimizations such as JIT and AOT compilation, allowing Wasm bytecode to be executed efficiently by utilizing the most suitable runtime for each environment.
The C/R mechanism saves the internal state of a VM executing Wasm bytecode within one runtime and restores it in another runtime on a different server, thereby enabling live migration of applications compiled into Wasm bytecode while preserving their execution state.

However, the diversity of runtime implementations and runtime-specific performance optimizations complicate cross-runtime C/R.
Although Wasm defines specifications for the resources of the VM created by a runtime, the actual resource implementations depend on each runtime's design, resulting in different representations of internal execution states.
Tailoring C/R mechanisms to different runtime implementations and converting internal states between them requires modifications to all runtimes in use, making cross-runtime C/R highly challenging and limiting the flexibility of combining the most suitable runtimes for each environment.
Moreover, runtimes employing performance optimizations such as JIT or AOT compilation translate Wasm bytecode into architecture-specific native binaries, which leads to internal execution states existing outside the VM itself.
Performing C/R in such runtimes requires handling CPU-architecture-specific states and conversions across architectures, which demands significant architecture-dependent implementations and undermines the architecture neutrality of Wasm.

In response to the above challenges, we propose a C/R mechanism using a self-hosted Wasm runtime, where the runtime itself is compiled into Wasm. In this approach, a lightweight Wasm runtime with C/R capability is compiled into Wasm bytecode and runs on any Wasm runtime.
By using the internal execution state of the self-hosted Wasm runtime, performing C/R eliminates the need for modifications to existing Wasm runtimes and for architecture-specific implementations.
Furthermore, to mitigate the performance degradation caused by duplicated runtime execution when using a self-hosted Wasm runtime, we designed a runtime and optimization techniques that assume compilation into Wasm from the outset.
To the best of our knowledge, this is the first Wasm runtime implementation specialized for self-hosting.

\noindent The contributions of this paper are as follows:
\begin{itemize}
  \item Proposal of a self-hosted Wasm runtime that achieves C/R in a manner neutral to both runtimes and performance optimizations.
  \item Design of a Wasm runtime specialized for self-hosting and comparison of its performance with existing self-hostable runtimes.
  \item Evaluation of the execution overhead introduced by the self-hosted Wasm runtime and discussion of its impact on runtime performance.
\end{itemize}

\section{Runtime-Neutral WebAssembly Checkpoint/Restore}
Inter-mingling appropriate Wasm runtimes suited to the characteristics and requirements of each edge server and cloud server can optimize the balance of performance, resource consumption, and functionality of applications while maintaining a unified binary representation.
Wasmtime provides high-performance bytecode execution through JIT compilation and is utilized in cloud platforms \cite{wasmtime-serverless}.
WebAssembly Micro Runtime (WAMR) is designed for embedded devices, supporting execution on device-oriented CPUs with low memory consumption \cite{wamr-low-memory-usage}.
WasmEdge offers extensions suited for executing AI workloads on edge servers.
Using Wasm runtimes tailored to each computing server allows applications to fully exploit the characteristics of each environment.

Application live migration based on C/R across heterogeneous Wasm runtimes optimized for each environment enables load balancing and service continuity that adapts to user mobility.
The migration targets are stateful applications such as in-memory databases, key–value stores, and stream processing, which maintain data internally and whose processing content or order depends on that data.
Edge–Edge live migration allows services to always provide low-latency processing by utilizing the edge server closest to the moving user \cite{edge-edge-live-migration}.
Edge–Cloud live migration enables offloading of computationally intensive tasks to high-performance servers \cite{wasm-edge-cloud-migration}.
Furthermore, prior work has shown that considering the characteristics of each runtime and switching runtimes according to the nature of the processing can improve application performance \cite{funeral-runtime-switching}.

\subsection{Challenges}
The internal execution state of the VM required for C/R varies significantly due to differences in runtime implementations and the performance optimization strategies implemented in each runtime, and thus, an approach to resolve these differences is required.
The execution state that changes during bytecode execution includes the value stack that stores numeric values, the frame stack that holds the states of executing functions, the linear memory that stores non-numeric data, local and global variables, and the program counter that indicates the position of the next bytecode instruction.
Although the Wasm specification defines these components, their implementations depend on each runtime.
Some Wasm runtimes use JIT or AOT compilation to translate bytecode into native binaries.
For example, they expand the value stack into registers, causing execution states to be scattered outside the VM, which makes it difficult to perform C/R based solely on the execution state of the VM.
Furthermore, differences in code generation granularity and applied optimizations make it challenging to apply C/R across different performance optimization strategies, such as JIT and AOT.

For live migration at the edge, it is also important to minimize downtime by reducing the internal execution state required for C/R and by enabling C/R to be performed at arbitrary points during execution.
Applications running on edge servers are often latency-sensitive \cite{edge-cr-latency-sensitive} or subject to user mobility, so if C/R points are restricted and live migration cannot be performed promptly, service quality may degrade.
In addition, edge servers have lower available computing power and network quality compared to cloud servers, and a large internal execution state for C/R increases the time required for both processing and transfer.
Therefore, performing C/R promptly with a minimal internal execution state at the necessary timing is essential for C/R at the edge.

\section{Limitations of Existing Approaches}
Existing Wasm C/R approaches have limitations in performing C/R across heterogeneous runtimes, in the size of the internal execution state, or in the ability to perform C/R at arbitrary points.

C/R approaches that modify runtimes are inherently difficult to implement in a manner neutral to heterogeneous runtimes and optimization strategies. In these approaches, each runtime must implement its own C/R mechanism and perform conversions between different internal state implementations \cite{funeral-runtime-switching}.
Furthermore, additional custom internal states need to be introduced for these conversions, and the complexity increases as the number of supported runtimes grows.
Such additional internal states may also increase the size of the internal execution state.
Another approach is to insert C/R-related native code into native binaries at JIT or AOT compilation time, but this depends on specific compiler infrastructures such as LLVM, making it difficult to apply to diverse runtimes \cite{wharf}.

The bytecode modification approach \cite{bytecode-modification-cr} inserts C/R code into Wasm bytecode before it is executed on a Wasm runtime.
Since the modified Wasm bytecode can be executed on any runtime, this approach achieves runtime and optimization neutrality.
However, since the points at which C/R should be performed cannot be determined in advance, C/R mechanisms must be inserted into every function.
As a result, this approach causes an increase in the size of the Wasm bytecode, unnecessary runtime overhead from creating redundant checkpoints, and a lack of flexibility in the points at which C/R can be performed.
\section{Proposed Method}
We propose a self-hosted runtime-based approach to achieve runtime- and optimization-neutral C/R without requiring modifications to existing runtimes.
In this approach, a self-hosted Wasm runtime with C/R capability is compiled into Wasm and executed on any host Wasm runtime, and Wasm applications are executed on top of the self-hosted runtime, thereby eliminating the need to modify existing runtimes.

\subsection{Overview}

\begin{figure}[t]
    \centering
    \includegraphics[width=0.92\linewidth]{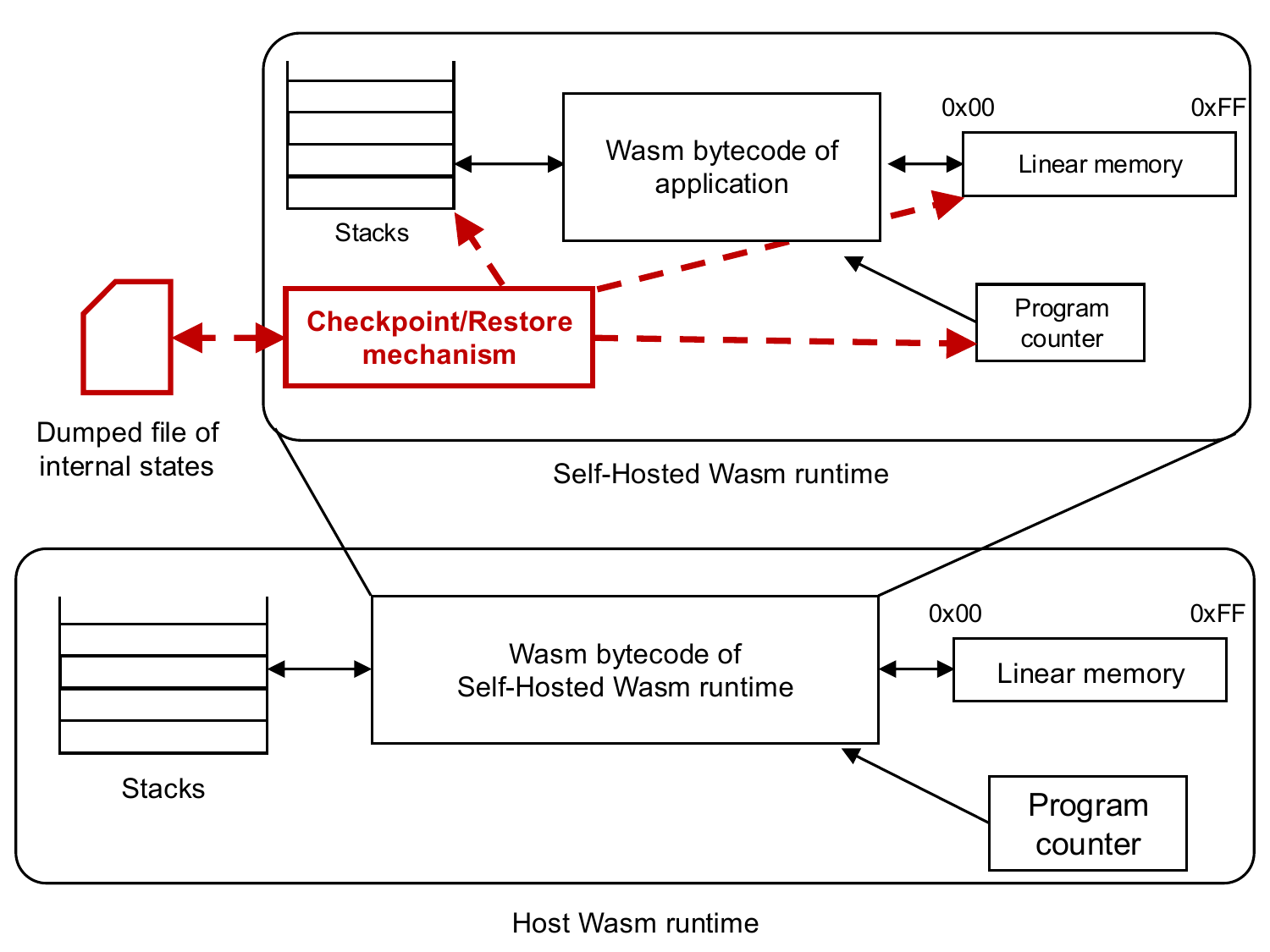}
    \caption{Overview of proposed approach.}
    \label{fig:architecture-overview}
\end{figure}

As shown in Fig.~\ref{fig:architecture-overview}, the proposed method executes Wasm bytecode on a host Wasm runtime through a self-hosted Wasm runtime.
The self-hosted Wasm runtime, similar to a conventional host runtime, creates a VM for executing Wasm bytecode and processes Wasm instructions.
The C/R mechanism of the proposed method checkpoints and restores the execution state of the VM created by the self-hosted Wasm runtime, including the program counter, the value and frame stacks, the linear memory, and local and global variables.
This approach enables C/R without requiring modifications to existing runtimes.

Our method not only eliminates the need to modify host Wasm runtimes but also removes the requirement to implement distinct C/R mechanisms for each optimization strategy.
Since Wasm bytecode is executed through the self-hosted Wasm runtime, the execution state of the running bytecode is entirely managed within the VM of the self-hosted runtime.
Although the performance optimizations of the host Wasm runtime are applied to the Wasm bytecode of the self-hosted runtime process, the internal execution state managed by the host runtime is translated.
Still, the execution state of the application’s Wasm bytecode managed by the self-hosted runtime remains unaffected.
Executing the self-hosted runtime on any host Wasm runtime enables C/R across heterogeneous runtimes and performance optimization strategies.

Furthermore, our approach checkpoints and restores specific internal execution states of the self-hosted VM, thereby reducing the size of the internal execution state compared with C/R of the entire OS process of the host Wasm runtime.

\subsection{Technical Challenges of Runtime Overhead}
Although our approach can achieve runtime- and optimization-neutral C/R, it faces the technical challenge of degraded performance in Wasm bytecode execution.
When Wasm bytecode is executed through the self-hosted Wasm runtime, instruction processing is performed in both the host and the self-hosted Wasm runtimes.
Since JIT and AOT compilations of the host Wasm runtime are applied only to the self-hosted Wasm runtime, the Wasm bytecode instructions executed by the self-hosted runtime cannot be optimized, making it difficult to improve application performance.

The duplicated execution of runtimes leads to an increase in the number of Wasm bytecode instructions executed on the host runtime and the duplication of computationally expensive processing.
Each instruction of the self-hosted Wasm runtime is compiled into a sequence of multiple Wasm bytecode instructions, which means that one application-level Wasm instruction is executed on the host runtime as instruction sequences for processing in the self-hosted runtime.
When a self-hosted Wasm runtime is used, the sandbox processing, which is one of the performance-intensive functions of Wasm runtimes, is executed separately in both the host and the self-hosted runtimes, resulting in increased performance overhead.
A Wasm runtime executes Wasm programs in an isolated sandbox to protect users and the execution environment from bugs or malicious programs.
The sandbox performs bytecode validation, such as type checks and value stack operation checks, before execution, and applies Software-based Fault Isolation (SFI), which performs boundary checks on linear memory during execution to ensure isolation.
SFI in Wasm is widely known as a major cause of significant performance degradation when applications are compiled into Wasm \cite{wasm-sfi}.
System interface implementations can also be performance-intensive in Wasm runtimes.

The WebAssembly System Interface (WASI) is a standard interface that allows Wasm bytecode to access host resources such as files, sockets, and environment variables.
In typical WASI implementations, buffers in the value stack or linear memory are utilized to invoke OS system calls or library functions, and the results are written back into linear memory, making the processing heavier than primitive Wasm instructions.
Therefore, when WASI calls are duplicated in both the self-hosted and the host runtimes, they can create a significant bottleneck.

\subsection{Optimization Techniques\\for Reducing Runtime Overhead}
We address the technical challenges by applying performance optimization techniques tailored for self-hosted Wasm runtimes.
To reduce the number of Wasm bytecode instructions executed on the host Wasm runtime, we minimize value stack operations.
We reduce duplicated sandboxing by omitting SFI and instruction validation in the self-hosted Wasm runtime.
We also minimize duplication of WASI implementations by enabling passthrough of WASI functions.

Value stack operations are frequently performed during Wasm bytecode execution, and optimizing these instructions can reduce the amount of instruction processing performed by the host runtime for the self-hosted Wasm runtime.
Since Wasm is a stack machine, operands (such as variables and constants) are pushed onto the value stack and popped when instructions are executed, after which the results are stored back on the value stack.
By allowing operands to be directly accessed without going through the value stack during instruction processing, the execution of value stack operation instructions can be minimized, thereby improving the overall instruction processing performance.

By creating a sandbox only in the host Wasm runtime, the responsibility for ensuring safety is delegated to the host runtime, which prevents unintended effects on the host environment while it can improve the execution performance of the self-hosted Wasm runtime.
The sandbox of the host runtime also secures the safety of the processing performed by the self-hosted Wasm runtime.
However, since sandboxing is not directly applied to the Wasm bytecode executed by the self-hosted runtime, malicious application behavior may not be trapped as long as the self-hosted runtime itself behaves normally, which may result in checkpointing and restoring corrupted internal states.
This represents a design trade-off in which runtime overhead is reduced by avoiding duplicate sandboxing, while sandbox-based security relies on the host runtime.

Simplifying the system interface implementation of the self-hosted Wasm runtime minimizes duplication of WASI implementations and reduces runtime overhead.
Since the self-hosted Wasm runtime is compiled into Wasm, it can call WASI functions implemented in the host Wasm runtime.
Thus, the WASI functions of the self-hosted runtime redirect arguments stored in the value stack or linear memory directly to the WASI functions implemented in the host runtime.
This enables passthrough of the WASI implementation in the self-hosted runtime.

\section{Implementation}
We implemented Chiwawa \footnote{Chiwawa: \url{https://github.com/oss-fun/chiwawa}}, a prototype of a self-hosted Wasm runtime.
Since it is an interpreter that executes Wasm bytecode sequentially, in principle, C/R can be performed at any point in Wasm bytecode execution.
Chiwawa consists of approximately 12,000 lines of Rust code and can be compiled only into Wasm.

Chiwawa reduces value stack operations by expanding instructions during parsing (before Wasm bytecode execution) based on dependencies between instructions.
For example, in numerical operations or memory access instructions, preceding instructions often push constants or local/global variables onto the value stack, and subsequent instructions sometimes store the results from the value stack into local or global variables.
By detecting such dependencies, operand information and destination information can be embedded directly into the instructions.
This allows instruction handlers to directly access and update operands without passing through the value stack, thereby significantly reducing value stack operations.

The WASI implementation of Chiwawa passes through application calls to WASI functions directly to those of the host Wasm runtime.
The Wasm bytecode executed in Chiwawa uses wasi-libc, a standard C library implementation designed for WASI environments.
Wasi-libc provides high-level functions such as printf, which internally invoke Chiwawa’s low-level WASI functions.
Instead of providing a general WASI function implementation, Chiwawa’s low-level WASI functions call the corresponding low-level WASI functions of the host Wasm runtime without intermediate buffer copies by using the linear memory addresses provided by Chiwawa.

The C/R mechanism of Chiwawa checkpoints and restores the data structures that manage the value stack, frame stack, linear memory, local and global variables, and the program counter.
The trigger for C/R is implemented by having Chiwawa use the multithreading extension to monitor the creation of a trigger file, which is currently being standardized and supported by many runtimes.
In principle, frequently checking for the existence of a file during Wasm bytecode execution is inefficient.
However, since WASI does not support OS signals and their standardization has not progressed, there is currently no efficient way to notify Chiwawa of events.
If OS signals are standardized in the future, a more efficient C/R trigger mechanism will become possible.

\section{Evaluation}
Using Chiwawa, we evaluated 1) the benefits of a runtime design tailored for self-hosting, 2) the performance overhead introduced by a self-hosted Wasm runtime, and 3) the benefits of self-hosted runtime-based C/R.
The experimental server is equipped with an Intel Xeon Silver 4208 processor (8 cores, 2.10GHz) and 32GB RAM, and runs Ubuntu 22.04.4 LTS (Linux 5.15.0).

\subsection{Performance Benefits of Design for Self-Hosting}
We evaluated the usefulness of a runtime design for self-hosting through performance comparisons between Chiwawa and Wizard \cite{wizard-instrumentation}, an existing self-hostable Wasm runtime.
Wizard is a state-of-the-art interpreter-based Wasm runtime with rich instrumentation capabilities for tracing, and can be used in a similar way to our work in that it can provide runtime-dependent functionality as Wasm bytecode through a self-hostable runtime.
For both Chiwawa and the self-hosted Wizard, the host runtime was Wizard.
This is because the self-hosted Wizard is compiled for the Wasm execution environment of an original programming language for research, and thus cannot be executed on Wasm runtimes other than Wizard.

To measure the execution performance of primitive Wasm instructions, we used two computationally intensive benchmarks: an n-body simulation and the calculation of pi based on the Leibniz series.
For the n-body simulation, the number of particles was set to 10,000, and for the Leibniz series, the number of terms was set to 1,000,000.
For each benchmark, we compared the average execution time and the average physical memory usage, measured as the Resident Set Size (RSS), over 50 runs.

Chiwawa was about 1.3 times faster than Wizard in the pi calculation and about 1.4 times faster in the n-body (see Table~\ref{table:wizard-execution-time}).
Regarding memory usage, Chiwawa achieved reductions of about 85\% and 71\%, respectively, compared to Wizard.

Chiwawa outperformed Wizard when self-hosted in terms of both execution performance and memory usage.
Wizard is not designed for self-hosting and therefore suffers from overheads such as duplicate sandboxing, which includes boundary checks of linear memory, and an increase in the number of executed instructions.
For example, WASI provides a security mechanism called preopen, in which the files accessible to Wasm bytecode are specified in advance, the sandbox opens them before execution, and only the corresponding file descriptors are made available to the Wasm bytecode.
In the case of Wizard, these are executed in both the host and the self-hosted runtime, causing duplicate file openings and file descriptor management, whereas Chiwawa, designed for self-hosting, eliminates such duplicate processing and also reduces the number of instructions executed in the host Wasm runtime.
This optimization not only reduces the number of instructions but also decreases the allocation and deallocation of stack slots, thereby achieving both improved execution performance and reduced memory usage.
These results demonstrate that, when using a self-hosted Wasm runtime to provide runtime-dependent functionality such as C/R in a runtime-independent manner, a design tailored for self-hosting is effective in reducing runtime overhead.

\begin{table}[t]
    \scalebox{1}{
    \begin{tabular}{cc}
        \begin{minipage}[t]{0.5\hsize}
            \subcaption{Execution time (sec.)}
            \label{table:wizard-execution-time}
            \centering
            \scalebox{0.89} {
                \begin{tabular}{l|r|r} \hline\hline
                    Benchmarks & Wizard & Chiwawa \\ \hline
                    pi-Leibniz & 26.791 & 19.620 \\
                    n-body & 14.554  & 9.932 \\ \hline
                \end{tabular}
            }
        \end{minipage}
        \begin{minipage}[t]{0.5\hsize}
            \subcaption{Memory usage (MB)}
            \label{table:wizard-memory-usage}
            \centering
            \scalebox{0.89} {
                \begin{tabular}{l|r|r} \hline\hline
                    Benchmarks & Wizard & Chiwawa \\ \hline
                    pi-Leibniz & 106.98 & 15.03 \\
                    n-body & 53.46  & 15.15 \\ \hline
                \end{tabular}
            }
        \end{minipage}
    \end{tabular}
    }
    \caption{Benchmark results.}
    \label{table:wizard-comparison}
\end{table}

\subsection{Performance Overhead of Self-Hosting}
\label{sec:performance-overhead-of-self-hosting}
To evaluate the runtime overhead introduced by a self-hosted Wasm runtime, we compared benchmark results using Chiwawa with results from executing Wasm bytecode directly on the host Wasm runtime.
We utilized wasmtime (version 32.0.0), which enables fast execution of Wasm bytecode through JIT compilation for host runtime for Chiwawa.
For the benchmark program, we used the SQLite3 benchmark\footnote{SQLite3 Benchmark: \url{https://github.com/ukontainer/sqlite-bench}}, which utilizes many WASI functions and is based on a lightweight DB widely used in edge computing.
In each benchmark, the number of entries was set to 100, the key size to 16 bytes, and the value size to 100 bytes.

\begin{figure}[t]
    \centering
    \includegraphics[width=1.0\linewidth]{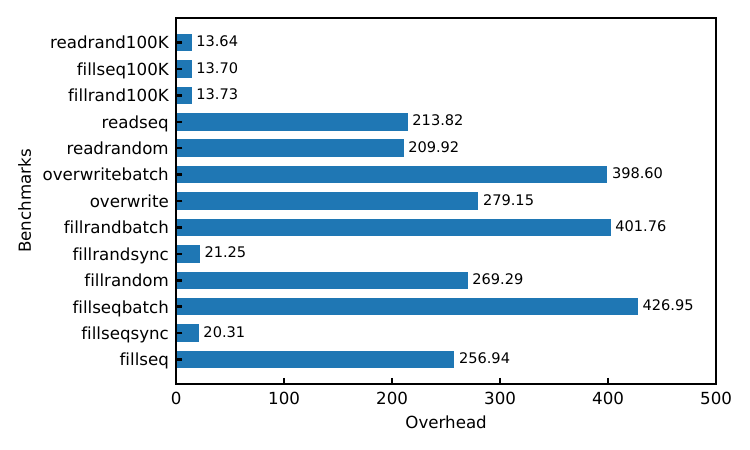}
    \caption{Relative execution time (normalized to wasmtime).}
    \label{fig:sqlite3}
\end{figure}

When using Chiwawa, the execution time increased in all benchmarks, ranging from about 13.64x at minimum to about 426.95x at maximum (see Fig.~\ref{fig:sqlite3}).
Regarding memory usage, the maximum during benchmark execution was about 77 MB with Chiwawa and about 33 MB when executed directly on wasmtime, representing an increase of about 2.33x.

While Chiwawa outperformed existing self-hostable Wasm runtimes, it still showed significant performance degradation compared to executing SQLite directly on the host Wasm runtime.
In particular, in the benchmark using SQLite's transaction functionality for batch processing, the overhead was outstanding, reaching about 400x.
Generally, batch processing improves performance by grouping multiple operations into a single transaction and thereby reducing the number of operations.
However, in the case of Chiwawa, the duplicated execution of runtimes exposed the overhead of storage engine processing, and as a result, the performance degradation became prominent. The overheads observed in other benchmarks also increased significantly for the same reason.
These results indicate that further reduction of Wasm bytecode instructions executed on the host Wasm runtime is essential.

\subsection{Benefits of Self-Hosted Runtime-Based C/R}
To evaluate the benefits of our proposed approach, we conducted a mock performance comparison with a runtime-modification-based C/R mechanism and a comparison of the size of the execution state dumped during C/R.
Wizard provides a high-performance instrumentation mechanism at the runtime level and, although it does not implement a C/R mechanism, it can monitor execution states equivalent to those required for C/R.
Therefore, it can be utilized for a mock evaluation against runtime-modification-based C/R.
The benchmarks were the same as those described in Section~\ref{sec:performance-overhead-of-self-hosting}.
Since Wizard is an optimized interpreter-based runtime, we used JIT-mode wasmtime and AOT-mode WAMR (v2.4.1) as the host runtimes for running Chiwawa.
For the comparison of dumped execution state sizes, we used Checkpoint/Restore In Userspace (CRIU), a C/R mechanism for Linux processes. By performing C/R on a Wasm runtime process, CRIU achieves runtime neutrality.
The Wasm runtimes used were wasmtime, interpreter/AOT-mode WAMR, and interpreter-mode WasmEdge (v0.15.0). The target program was Wasm bytecode running SQLite, and the checkpoint was taken after inserting 1,000 entries of data.

\begin{figure}[t]
    \centering
    \includegraphics[width=1.0\linewidth]{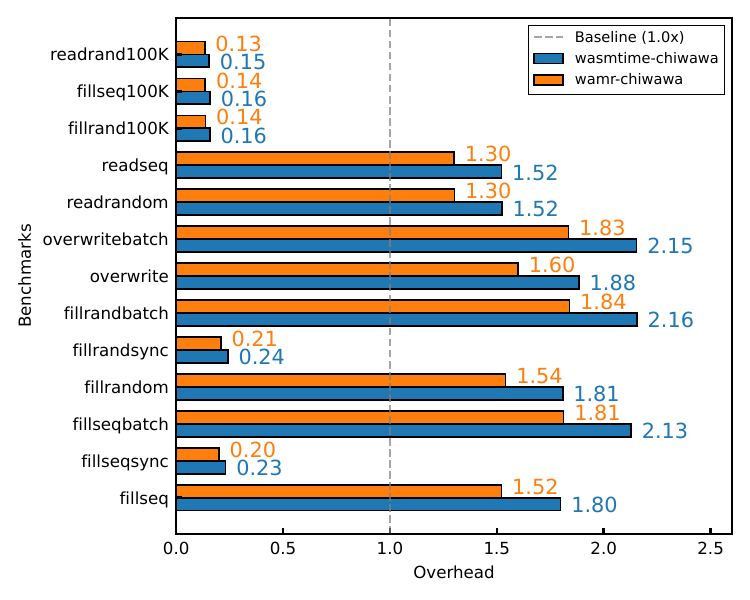}
    \caption{Relative execution time (normalized to wizard).}
    \label{fig:sqlite-wizard}
\end{figure}

Using wasmtime and WAMR as host runtimes, Chiwawa showed a relative execution time ranging from 0.13x at best to 2.16x at worst compared to Wizard (see Fig.~\ref{fig:sqlite-wizard}).
Although Chiwawa performs C/R through duplicated runtime execution, in some benchmarks, it achieved better performance than directly tracing the internal execution state of Wasm bytecode with Wizard.
Even at worst, the overhead was only 2.16x, which is significantly smaller than the runtime overhead evaluated in Section~\ref{sec:performance-overhead-of-self-hosting}.
This suggests that, even in runtimes such as Wizard that implement performance optimization techniques, tracing internal execution states is highly costly, and runtime-modification-based C/R may cause performance degradation comparable to duplicated runtime execution.
Furthermore, running Chiwawa on runtimes with performance optimizations such as JIT or AOT compilation can reduce the runtime overhead even further.

Chiwawa not only significantly reduced the size of the execution state compared to CRIU across all runtimes but also kept the sizes consistent across different runtimes (see Table \ref{table:cr-size}).
This is because C/R in Chiwawa targets only the parts of the execution state managed by the Wasm runtime that change during the execution of Wasm bytecode.
If CRIU were to achieve a similar reduction, it would be necessary to identify the Wasm bytecode state within the process execution state and provide a highly complex subset mechanism to checkpoint and restore only those parts.
Our approach ensures that, even if the internal implementations of the runtimes differ, no additional information is required to compensate for differences in state, and the size of the execution state can be consistently kept small.

\begin{table}[t]
    \caption{Checkpointed Wasm-bytecode state size (KB).}
    \label{table:cr-size}
    \centering
    \scalebox{1} {
        \begin{tabular}{l|r|r} \hline\hline
            Host runtimes & Chiwawa & CRIU \\ \hline
            Wasmtime & 1076 & 9232\\
            WAMR (Interpreter) & 1076 & 654484 \\
            WAMR (AOT) & 1076 & 14360 \\
            WasmEdge & 1076  & 89784 \\ \hline
        \end{tabular}
    }
\end{table}

\section{Conclusion}
We proposed a runtime-neutral C/R approach that utilizes the internal execution state managed by a self-hosted Wasm runtime that can be executed on any Wasm runtime and implemented Chiwawa, an optimized runtime designed to be self-hosted.
Experimental results showed that Chiwawa can eliminate runtime differences in C/R and enable efficient C/R with a minimal execution state.
However, while Chiwawa achieved a higher execution performance than making an existing Wasm runtime self-hostable, it still suffers from significant runtime overhead compared to execution without self-hosting.

In future work, we plan to develop optimization techniques for self-hosted Wasm runtimes that can further reduce runtime overhead, as well as C/R for Wasm bytecode that uses OS resources outside the Wasm runtime.
Currently, Chiwawa is implemented as a stack machine, but duplicated stack machines are inefficient, and there remains room for optimization in this part.
Furthermore, the C/R mechanism of Chiwawa cannot yet support applications that utilize OS-dependent resources such as sockets and files through WASI.
We will investigate approaches that are neutral to both runtimes and OSes.

\bibliographystyle{ACM-Reference-Format}
\bibliography{bib}

\end{document}